\documentclass[aps,noshowpacs,superscriptaddress,floatfix]{revtex4}

\usepackage{array}

\usepackage[utf8]{inputenc}
\usepackage{braket}
\usepackage{amsmath,amssymb,graphicx}
\usepackage{caption}
\usepackage{subcaption}
\usepackage{qcircuit}
\usepackage{xcolor}
\usepackage{verbatim}
\usepackage[normalem]{ulem}
\usepackage{graphicx}
\usepackage[rightcaption]{sidecap}

\newcommand{\R}{\mathbb{R}}

\def\ket#1{\mathinner{|{#1}\rangle}}
\usepackage{float}
\usepackage{multirow}
\usepackage[ruled,vlined]{algorithm2e}
\usepackage{amsmath}

\usepackage{hyperref}

\begin{document}
\title{Improved clinical data imputation via classical and quantum determinantal point processes}

\author{Skander Kazdaghli}
\affiliation{QC Ware, Palo Alto, USA and Paris, France}

\author{Iordanis Kerenidis}
\affiliation{Université de Paris, CNRS, IRIF, 8 Place Aurélie Nemours, Paris 75013, France}
\affiliation{QC Ware, Palo Alto, USA and Paris, France}

\author{Jens Kieckbusch}
\affiliation{Emerging Innovations Unit, Discovery Sciences, BioPharmaceuticals R\&D, AstraZeneca, Cambridge, UK}

\author{Philip Teare}
\affiliation{Centre for AI, Data Science \& AI, BioPharmaceuticals R\&D, AstraZeneca, Cambridge, UK}


\begin{abstract}
    Imputing data is a critical issue for machine learning practitioners, including in the life sciences domain, where missing clinical data is a typical situation and the reliability of the imputation is of great importance. Currently, there is no canonical approach for imputation of clinical data and widely used algorithms introduce variance in the downstream classification.
    Here we propose novel imputation methods based on determinantal point processes that enhance popular techniques such as the Multivariate Imputation by Chained Equations (MICE) and MissForest. Their advantages are two-fold: improving the quality of the imputed data demonstrated by increased accuracy of the downstream classification; and providing deterministic and reliable imputations that remove the variance from the classification results.
    We experimentally demonstrate the advantages of our methods by performing extensive imputations on synthetic and real clinical data.
    We also perform quantum hardware experiments by applying the quantum circuits for DPP sampling, since such quantum algorithms provide a computational advantage with respect to classical ones. We demonstrate competitive results with up to ten qubits for small-scale imputation tasks on a state-of-the-art IBM quantum processor. 
    Our classical and quantum methods improve the effectiveness and robustness of clinical data prediction modeling by providing better and more reliable data imputations. These improvements can add significant value in settings demanding high precision, such as in pharmaceutical drug trials where our approach can provide higher confidence in the predictions made.
\end{abstract}

\maketitle

Missing data is a recurring problem in machine learning and in particular for clinical datasets, where it is common that numerous feature values are not present for reasons including incomplete data collection, discrepancies in data formats and data corruption \cite{Luo21, Emmanuel2021, alma17, william2000}. Machine learning is routinely used in life science and clinical research for prediction tasks, such as diagnostics \cite{qin2019machine} and prognostics \cite{booth2021development}, as well as estimation tasks, such as biomarker proxies \cite{wang2017predicting} and digital biomarkers \cite{rendleman2019machine}. Beyond the research setting, machine learning is becoming more commonplace as regulated Software as a Medical Device, where machine learning models are influencing - or making - clinical decisions that affect patient care. 

Machine learning algorithms typically require complete data sets and missing values can significantly affect the quality of the machine learning models trained on such data. This is in large part due to the fact that there can be different underlying reasons for the missingness: for example, feature values can be missing completely at random (MCAR), missing at random (MAR) and missing not at random (MNAR), each one with their own characteristics. 

Despite its importance for clinical trials, there is no canonical approach for dealing with missingness and finding appropriate, effective and reproducible methods remains a challenge.
A common way to deal with missing clinical data is to exclude subjects that do not have the complete set of feature values present. A drawback of this approach is that excluding subjects can in fact introduce significant biases in the final model. For example, it can result in the model being trained to be more effective for the type of subjects that are likely to have complete data than for those that do not. Moreover, the effectiveness and reliability of clinical trials is reduced when subjects with missing feature values are excluded from the clinical trial.  

\begin{figure}[!h]
    \centering
\includegraphics[width=0.9\linewidth]{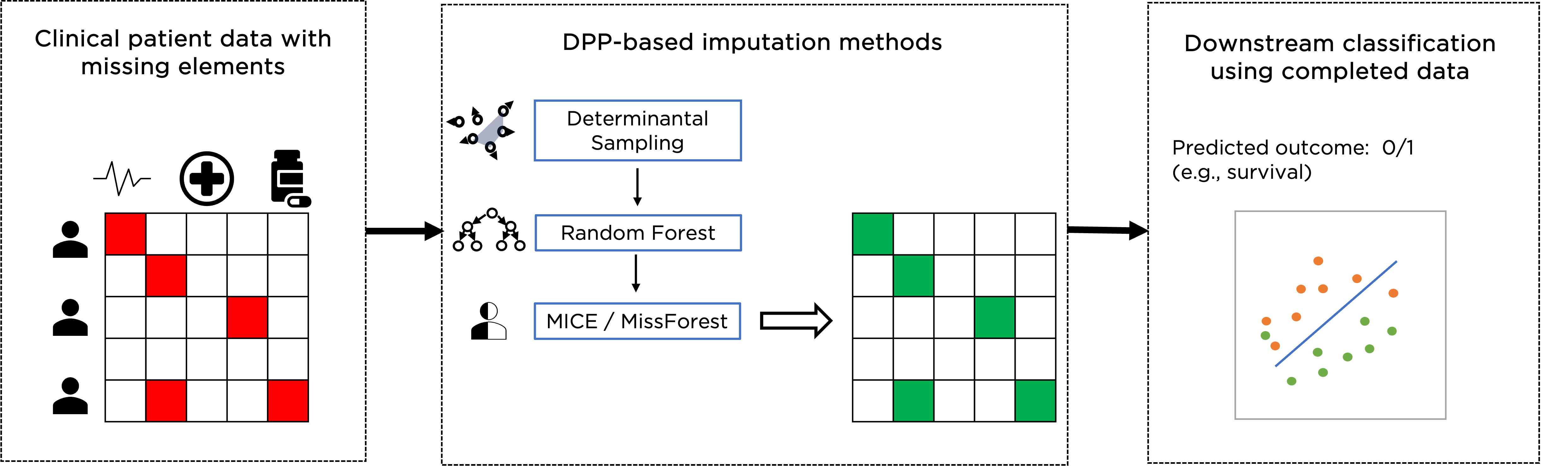}
    \caption{Example of overall workflow for patient management through clinical data imputation and downstream classification}
    \label{fig:workflow}
\end{figure}

Data imputation is an alternative to the complete dataset approach, where subjects with missing feature values are not excluded. Instead, missing values are imputed to create a complete dataset that is then used for a classification task as shown in Fig. \ref{fig:workflow}. There are different ways to achieve this, including “filling” missing values with zeros, or with the mean value of the feature across all subjects that have such a value present. These methods provide consistent imputation results, but there are important caveats for using such simple methods, since they ignore possible correlations between features and can make the dataset appear more homogeneous than it really is. 
More advanced data imputation methods have been proposed in the literature: iterative methods include the multivariate imputation by chained equations (MICE) \cite{mice11} and MissForest \cite{mf11} algorithms, and deep learning methods include  GAIN (generative adversarial imputation nets) \cite{gain} and MIWAE (missing data importance-weighted autoencoder) \cite{miwae}. 
Recent results \cite{AZ22} have shown that for clinical data two iterative imputation methods, MiceRanger, that uses predictive mean matching, and MissForest, that uses Random Forests to predict the missing values of each feature using the other features, provide the best results and have been used here as a baseline.

\begin{figure}
    \centering
    \includegraphics[width=0.65\linewidth]{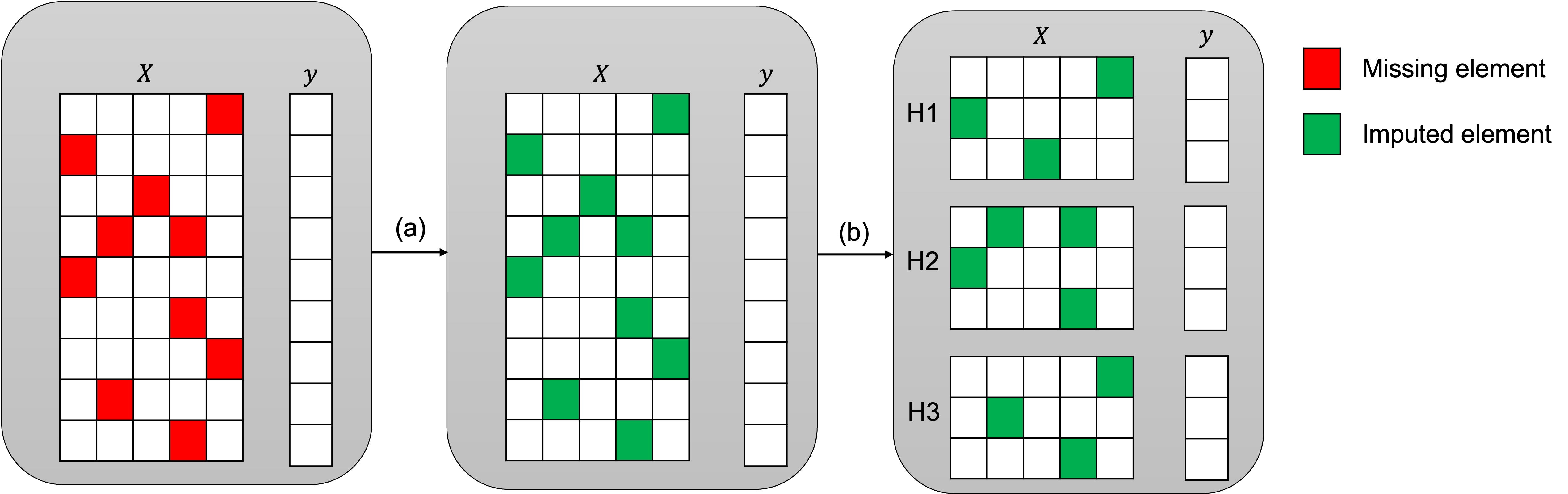}
    \caption{Imputation and downstream  classification procedure to benchmark the imputation method's performance. First, the imputer is trained on the whole observed dataset X as shown in step (a). In step (b), the imputed data is split into 3 consecutive folds (holdout sets H1, H2 and H3) then a classifier is trained on each combination of 2 holdout sets (development sets D1, D2 and D3) and the AUC is calculated for each holdout set.}
    \label{fig:training_procedure}
\end{figure}

Several metrics are routinely used to quantify the quality of data imputation:
point-wise discrepancy measures include root mean square error (RMSE), mean absolute error (MAE) and coefficient of determination ($R^2$). Feature-wise discrepancy measures include Kullback-Leibler divergence, two-sample Kolmogorov-Smirnov statistic or (2-)Wasserstein distance. Ultimately, the quality and reliability of imputations can be measured by the performance of a downstream predictor, which is usually the AUC (area under the receiver operating curve) for a classification task. In practical terms, the performance of the downstream classifier is usually of highest importance for clinical data sets: for example, in one of our datasets, the classifier denominates a binary outcome of a critical care unit stay (e.g. survival) for each patient. Accordingly, we have used AUC for the classification task here on different holdout sets (see Fig. \ref{fig:training_procedure}) to assess the performance of our novel methods.

In order to increase the resulting AUC, we combine the MissForest and MiceRanger imputation methods with determinantal sampling, based on determinantal point processes (DPP) \cite{KT11, derezinski2021determinantal} which favors samples that are diverse and thus reduces the variance of the training of each decision tree, which in turn provides more accurate models. In essence, determinantal sampling picks subsets of data according to a distribution that gives more weight to subsets of data that contain diverse datapoints. More precisely, each subset of datapoints is picked according to the volume encapsulated by these datapoints. The determinantal distribution increases the attention given to uncommon or out-of-the-ordinary data points rather than biasing the learning process towards the more commonly found data, which can improve the overall prediction accuracy in particular for unbalanced datasets as is often the case for clinical data \cite{DW20}. Determinantal sampling for regression and classification tasks with full data has been proposed previously for linear regressors \cite{D18} and for Random Forest training for a financial data classification use case where it outperformed the standard Random Forest model \cite{itau23}. However, an inherent feature of standard Random Forest and determinantal sampling algorithms is randomness that produces data imputations that vary from one run of the algorithm to the next. This is often undesirable, since the downstream classification performance can also be affected, which motivated us to apply a deterministic version of determinantal sampling \cite{SF21} within the Random Forests of the imputation methods to provide more robust and reliable imputations.

 Through deterministic determinantal sampling we address two challenges in data imputation: first, we provide improved data imputation methods that can increase the performance of the downstream classifier; and second, we remove the variance of the common stochastic and multiple imputation methods, thus ensuring reproducibility, easier integration in machine learning workflows, and compliance with healthcare regulations. While these improvements are of particular relevance for clinical data, our algorithms can also be advantageous for other imputation tasks where improving downstream classification and removing variance is of importance.
 
 In order to demonstrate this improvement, we apply our methods to two classification datasets: a synthetic dataset and a public clinical dataset where the predicted outcome is the survival of the patient.
 
In addition, we explore the potential of quantum computing to speed up these novel imputation methods: we provide a quantum circuit implementation of the determinantal sampling algorithm that offers a computational advantage compared to its classical counterpart. The best classical algorithms for determinantal sampling take in practice cubic time in the number of features to provide a sample \cite{DW20}. 
In contrast, the quantum algorithm we present here, based on theoretical analysis in \cite{KP22}, has running time that scales linearly  with the number of features. We measure running time as the depth of the necessary quantum circuits, given that the quantum processing units that are being developed currently offer the possibility of performing parallel operations on disjoint qubits.

This suggests that with the advent of next generation quantum computers with more and better qubits, one could also expect a computational speedup in performing determinantal sampling using a quantum computer. Here, we demonstrate competitive results with up to ten qubits for small-scale imputation tasks on a state-of-the-art IBM quantum processor.

This work combines classical \cite{DW20}, \cite{SF21} and quantum \cite{KP22} DPP algorithms with widely used data imputation methods, resulting in novel data imputation algorithms that can improve performance on classical computers while also having the potential of a quantum speedup in the future.






\section*{RESULTS}


We provide in Methods a detailed description of our four imputation methods, DPP-MICE, DPP-MissForest, detDPP-MICE and detDPP-MissForest. All of them are based on iterative imputation methods that use the observed values of every column to predict the missing values. The model used to fill missing values in each column is the Random Forest classifier. 
Our imputation methods replace the standard Random Forest used by the original miceRanger and MissForest imputers by the DPP-Random Forest model, for our first two imputers, and the detDPP-Random Forest for the latter two. The DPP-Random Forest model subsamples the data for each decision tree using determinantal sampling instead of uniform sampling, while the detDPP-Random Forest model deterministically picks for each decision tree the subset of data that has the maximum probability according to the determinantal distribution. 
We also demonstrate a computationally advantageous way to perform the determinantal sampling on quantum computers. 

In order to benchmark the different imputation methods, we used two types of datasets with a categorical outcome variable. First, a synthetic dataset, created using the scikit-learn method \emph{make\_classification}. It consists of 2000 rows with 25 informative features. This is useful to study the imputation quality where features have equal importance. Second, the MIMIC-III dataset \cite{MIMIC_Ref}: The Medical Information Mart for Intensive Care (MIMIC) dataset which is a freely available clinical database. It is comprised of data for patients who stayed in critical care units at the Beth Israel Deaconess Medical Center between 2001 and 2012. It contains the data of 7214 patients with 14 features.

We also applied two types of missingness on these datasets: MCAR (missing completely at random), where the missingness distribution is independent of any observed or unobserved variable; and MNAR (missing not at random), where the missingness distribution depends on the outcome variable. We expect similar results to hold for the MAR case as well, but it was not considered in this work.

We present the numerical results in terms of the AUC of the downstream classification task in Table \ref{tab:results} and provide graphs of the results in tables \ref{tab:mice_sim} and \ref{tab:mf_sim}. Each experiment was run ten times with different random seeds to get the variance of the results. 

Overall, DPP-MICE and DPP-MissForest provide improved results compared to their classical baseline MICE and MissForest. This is the case for both the synthetic and the MIMIC datasets and for both MCAR and MNAR missingness.
Even more interestingly, the detDPP-MICE and detDPP-MissForest collapse the variance of the imputed data to 0 and moreover lead in most cases to even higher AUC than the expectation of the previous methods.

\begin{table}[!ht]
    \centering
    \begin{tabular}{|c|c|c|c|c|c||c|c|c|}
    \hline
        Dataset & Missingness & Set & MICE & DPP-MICE & detDPP-MICE & MissForest & DPP-MissForest & detDPP-MissForest  \\ \hline
        \multirow{6}{*}{SYNTH} & \multirow{3}{*}{MCAR} & H1 & 0.8318±0.0113 & \textbf{0.835±0.0083} & \textbf{\underline{0.8352}} & 0.8525±0.0044 & \textbf{0.8552±0.0049} & \textbf{\underline{0.8582}}  \\ \cline{3-9}
        & ~ & H2 & 0.8316±0.008 & \textbf{0.8369±0.0128} & \textbf{\underline{0.84}} & 0.8465±0.0057 & \textbf{0.849±0.003} & \textbf{\underline{0.8491}}  \\  \cline{3-9}
        & ~ & H3 & 0.8205±0.0127 & \textbf{0.8266±0.0096} & \textbf{\underline{0.8272}} & 0.8436±0.0031 & \textbf{0.8452±0.0048} & \textbf{\underline{0.855}}  \\  \cline{2-9}
        ~ & \multirow{3}{*}{MNAR} & H1 & 0.8903±0.0046 & \textbf{0.8915±0.007} & \textbf{\underline{0.8934}} & 0.7133±0.0063 & \textbf{0.7171±0.01} & \textbf{\underline{0.7185}}  \\  \cline{3-9}
        & ~ & H2 & 0.8755±0.01 & 0.8745±0.0072 & \textbf{\underline{0.8955}} & 0.7052±0.0036 & \textbf{0.7124±0.0078} & \textbf{\underline{0.7167}}  \\  \cline{3-9}
        ~ & ~ & H3 & 0.9003±0.0059 & \textbf{0.9005±0.006} & 
        \textbf{\underline{0.9041}} & 0.769±0.0103 & \textbf{0.7773±0.0129} & \textbf{\underline{0.7905}}  \\  \hline
        \multirow{6}{*}{MIMIC} & \multirow{3}{*}{MCAR} & H1 & 0.7621±0.0046 & \textbf{0.7628±0.0049} & \textbf{\underline{0.7641}} & 0.7687±0.0012 & \textbf{0.77±0.0013} & \textbf{\underline{0.771}}  \\  \cline{3-9}
        ~ & ~ & H2 & 0.7541±0.0037 & 0.7532±0.0047 & \textbf{\underline{0.7619}} & 0.7649±0.0019 & \textbf{\underline{0.777±0.0019}} & \textbf{0.7707}  \\  \cline{3-9}
        ~ & ~ & H3 & 0.7365±0.0055 & \textbf{0.7394±0.0052} &  \textbf{\underline{0.7471}} & 0.7485±0.001 & \textbf{0.7507±0.0017} & \textbf{\underline{0.7515}}  \\  \cline{2-9}
        ~ & \multirow{3}{*}{MNAR} & H1 & 0.77±0.0026 & \textbf{0.7717±0.0036} & \textbf{\underline{0.7722}} & 0.6616±0.0065 & \textbf{0.6715±0.07} & \textbf{\underline{0.6760}}  \\  \cline{3-9}
        ~ & ~ & H2 & 0.777±0.0064 & \textbf{\underline{0.7818±0.0029}} & \textbf{0.7812} & 0.6748±0.0045 & \textbf{0.6778±0.0048} & \textbf{\underline{0.6798}}  \\  \cline{3-9}
        ~ & ~ & H3 & 0.7324±0.0047 & \textbf{0.7363±0.0031} & \textbf{\underline{0.7403}} & 0.6368±0.0034 & \textbf{0.64±0.004} & \textbf{\underline{0.6419}} \\ \hline
    \end{tabular}
     \caption{AUC results for the SYNTH and MIMIC-III datasets, with MCAR and MNAR missingness, three holdout sets, and six different imputation methods. Values are expressed as mean ± SD (standard deviation) of 10 values for each experiment. DPP-MICE and detDPP-MICE are bold when outperforming MICE and the underlined one is the best of the three. DPP-MissForest and detDPP-MissForest are bold when outperforming MissForest and the underlined one is the best of the three.}
     \label{tab:results}
\end{table}

\subsubsection*{DPP-MICE, and detDPP-MICE outperform MICE}

We present the performance results of MICE-based methods in terms of the AUC of the downstream classification task using an XGBoost classifier, which has been shown to be the strongest classifier for such datasets \cite{AZ22}. We used the default parameters of the classifier, since our focus is comparing the different imputation methods.
In each case, the original dataset with induced missing values is imputed using MICE, DPP-MICE or detDPP-MICE, then it is divided into 3 folds of Development/Holdout sets.
The downstream classifier is then trained on each development set and its performance is measured by the AUC for the corresponding holdout set. The results appear in Table \ref{tab:results} and in the figures in Table \ref{tab:mice_sim}.

The imputation procedure is performed for a total of 10 iterations over all the columns and for each column, a (DPP) Random Forest regressor is trained using 10 trees.
For each Random Forest training, the dataset is divided into batches of 150 points each and DPPs are used to sample from every batch.

\begin{table}[!h]
    \centering
    \begin{tabular}{c|c|c}
        &MCAR & MNAR\\
        \hline 
         \rotatebox[y=2.5cm]{90}{SYNTH}&\includegraphics[width=6cm]{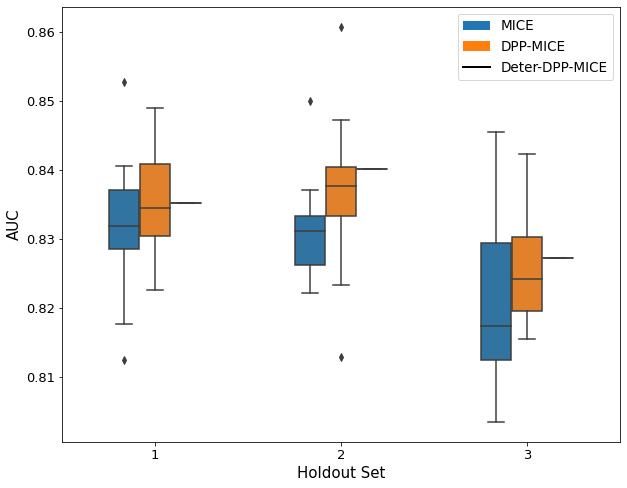}& \includegraphics[width=6cm]{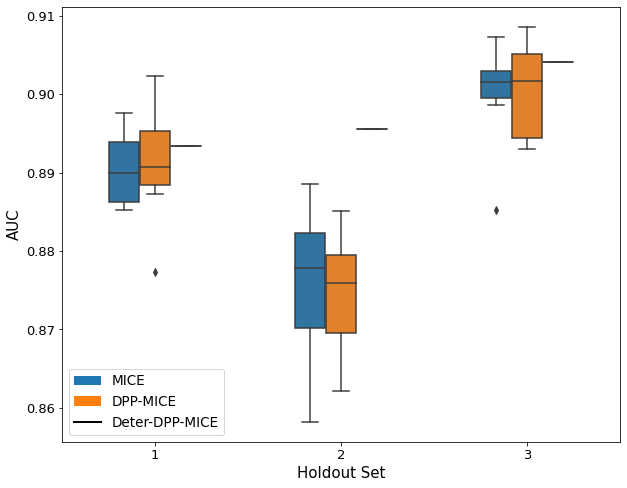} \\
        \hline 
         \rotatebox[y=2.5cm]{90}{MIMIC}&\includegraphics[width=6cm]{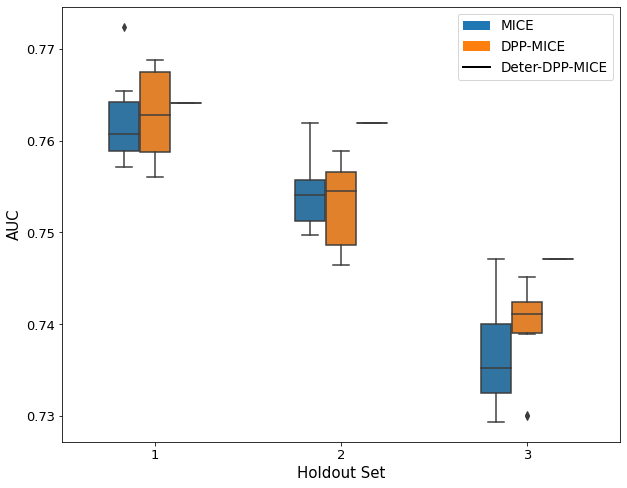}&  \includegraphics[width=6cm]{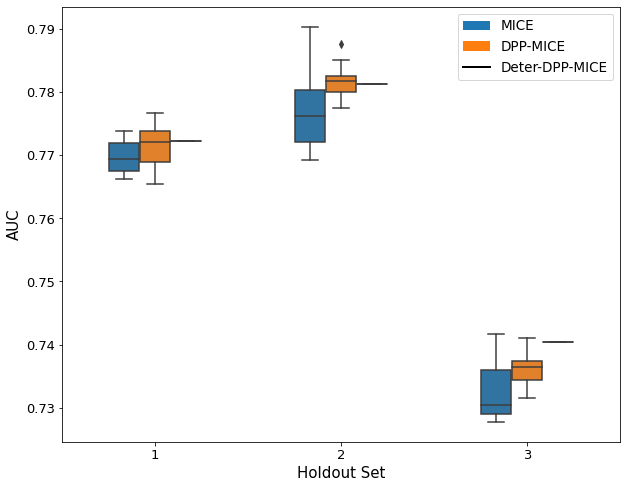}\\
         & 
    \end{tabular}
    \caption{AUC results on the different holdout sets after imputation using MICE, DPP-MICE and detDPP-MICE. In the case of MICE and DPP-MICE, the boxplots correspond to 10 AUC values for 10 iterations of the same imputation and classification algorithms, depicting the lower and upper quartiles as well as the median of these 10 values. The AUC values are the same for every iteration of the detDPP-MICE algorithm.}
    \label{tab:mice_sim}
\end{table}

The results show that across the twelve in total dataset experiments DPP-MICE outperforms MICE on expectation in ten of them, while detDPP-MICE provides a single deterministic imputation which outperforms the expected result from MICE in all twelve datasets and from DPP-MICE eleven out of twelve times.

\subsubsection*{DPP-MissForest and detDPP-MissForest outperform MissForest}

Here we present the performance results of MissForest-based methods in terms of the AUC of the downstream classification task using again an XGBoost classifier. 
In each case, the original dataset with induced missing values is imputed using MissForest, DPP-MissForest or detDPP-MissForest, then it is divided into 3 folds of Development/Holdout sets. The downstream classifier is again then trained on each development set and its performance is measured by the AUC for the corresponding holdout set. The results appear in Table \ref{tab:results} and in the figures in Table \ref{tab:mf_sim}. The specifics of the Random Forest training are the same as in the case of MICE.

\begin{table*}[!h]
    \centering
    \begin{tabular}{c|c|c}
        &MCAR & MNAR \\
        \hline 
         \rotatebox[y=2.5cm]{90}{SYNTH}&\includegraphics[width=6cm]{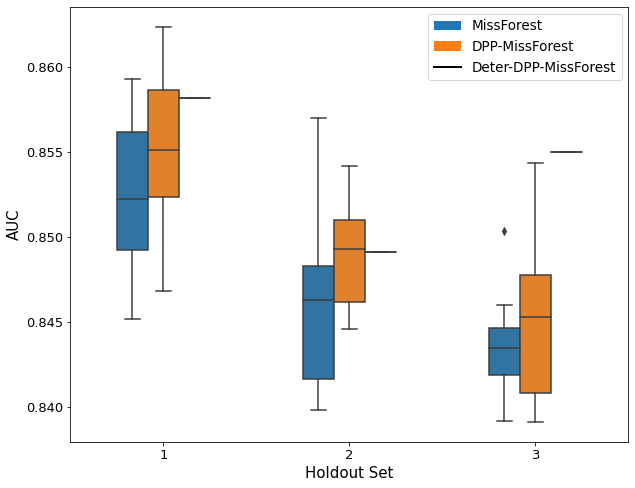}& \includegraphics[width=6cm]{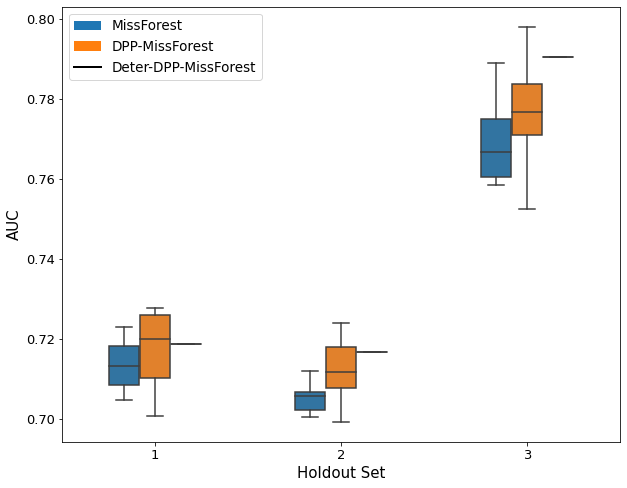} \\
        \hline 
         \rotatebox[y=2.5cm]{90}{MIMIC} &\includegraphics[width=6cm]{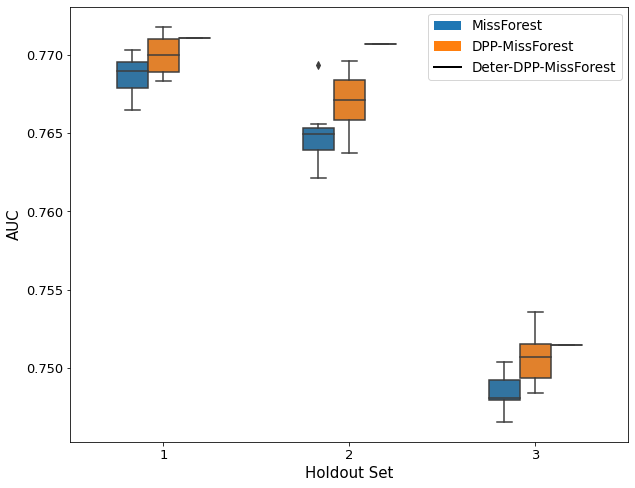}&  \includegraphics[width=6cm]{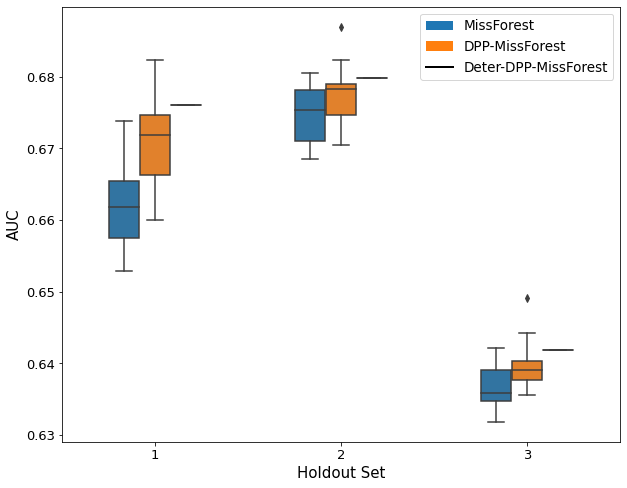}\\
         & 
    \end{tabular}
    \caption{AUC results on the different holdout sets after imputation using MissForest, DPP-MissForest and detDPP-MissForest. In the case of MissForest and DPP-MissForest, the boxplots correspond to 10 AUC values for 10 iterations of the same imputation and classification algorithm, depicting the lower and upper quartiles as well as the median of these 10 values. The AUC values are always the same for every iteration of the detDPP-MissForest algorithm.}
    \label{tab:mf_sim}
\end{table*}

The results show that across all experiments, DPP-MissForest outperforms MissForest in all twelve of them, while detDPP-MissForest provides a single deterministic imputation which outperforms the expected result from MissForest in all twelve datasets and from DPP-MissForest in eleven out of twelve times.

\subsubsection*{Quantum hardware implementation of DPP-MissForest results in competitive downstream classification}

As we describe in Methods, quantum computers can in principle be used to offer a computational advantage in determinantal sampling. In order to better understand the state-of-the-art of current quantum hardware, we used a currently available quantum computer to perform determinantal sampling within a DPP-MissForest imputation method for scaled-down versions of the synthetic and MIMIC datasets.

\begin{itemize}
    \item Reduced synthetic dataset: 100 points and 3 features, created using the sklearn method \emph{make\_classification}.
    
    \item Reduced MIMIC dataset: 200 points and 3 features. The three features were chosen from the original dataset features based on low degree of missingness and their predictiveness of the downstream classifier and they were: “Oxygen saturation std”, “Oxygen saturation mean” and “Diastolic blood pressure mean”.
\end{itemize}

\begin{figure}[!h]
    \centering
    \includegraphics[width=0.4\linewidth]{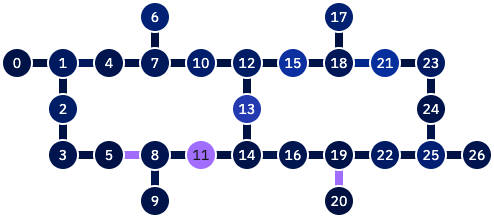}
    \caption{IBM Hanoi 27-qubit quantum processor}
    \label{fig:hanoi}
\end{figure}

For the purposes of our experiments, we used the “ibm\_hanoi” 27-qubit quantum processor shown in Fig. \ref{fig:hanoi}. We implemented quantum circuits with up to 10 qubits. We also performed quantum simulations using the qiskit noiseless simulator. The decision trees of the DPP-Random Forests used by the imputation models are trained using batches of decreasing sizes (see Table \ref{tab:hw_sizes}). For example, for the algorithm with batch size equal to 10, the algorithm first samples two out of the ten data points to use for the first decision tree, then from the remaining eight datapoints it picks another two for the second tree, then two from the remaining six, and last two from the remaining four. In other words, we train four different trees, and each time we use a quantum circuit with number of qubits equal to 10, 8, 6, and 4, to perform the respective determinantal sampling.

\begin{table}[!h]
    \centering
    \begin{tabular}{c|c|c|c|c}
        Batch Size & Tree 1 & Tree 2 & Tree 3 & Tree 4\\\hline
        7 & (7,2) & (5,2) &-&-\\\hline
        8 & (8,2) & (6,2) & (4,2) & -\\\hline
        10 & (10,2) & (8,2) & (6,2) &(4,2)
    \end{tabular}
    \caption{Data matrix sizes used by the quantum DPP circuits to train each tree. The number of rows corresponds to the number of data points and is equal to the number of qubits of every circuit.}
    \label{tab:hw_sizes}
\end{table}

In the figures of Table \ref{tab:hw_fig} and in Table \ref{tab:hw} we provide for the different dataset experiments the AUC for MissForest, the simulated results of the quantum version of DPP-MissForest, and the actual hardware experimental results of running the quantum version of DPP-MissForest. Even for these very small datasets, when simulating the quantum version of DPP-MissForest, we demonstrate an increase in the AUC compared to the MissForest algorithm. 
This further highlights the potential advantages of determinantal sampling within imputation methods. Of note, running our algorithms on current hardware introduces variance in the downstream classifier. Importantly, this variance is due to noise in the hardware rather than inherent to the algorithm.


Our quantum hardware results are competitive with standard methods and in many cases close to the values expected from the simulation. In some cases, we observed a clear deterioration of the AUC due to the noise and errors in the quantum hardware. The results are closer to the simulations when using MCAR missingness with larger batch sizes that use more trees both for synthetic and the MIMIC datasets. As explained above, even though the algorithm with batch size 10 means using a quantum circuit with 10 qubits, the fact that we use four trees overall with a decreasing number of datapoints each time, and thus a decreasing number of qubits (namely, 10, 8, 6, and 4), results in an overall more reliable imputation.

\begin{table}[!h]
    \centering
    \begin{tabular}{c|c|c|c}
        &Batch size: 7 & Batch size: 8 & Batch size: 10\\
         &Number of trees: 2 & Number of trees: 3 & Number of trees: 4\\
        \hline 
        \rotatebox[y=1.5cm]{90}{MCAR SYNTH}
        &\includegraphics[width=5.75cm]{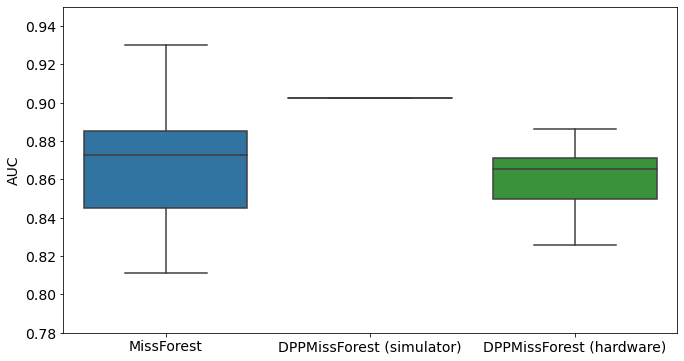}&\includegraphics[width=5.75cm]{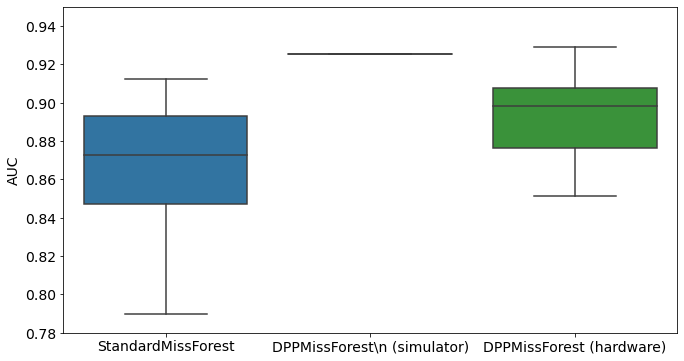}&\includegraphics[width=5.75cm]{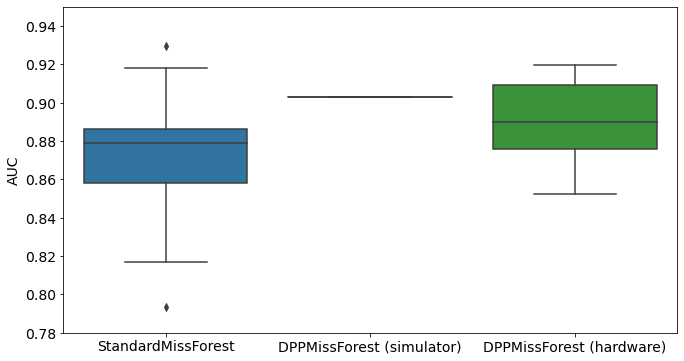}  \\
        \hline 
        \rotatebox[y=1.5cm]{90}{MCAR MIMIC}&\includegraphics[width=5.75cm]{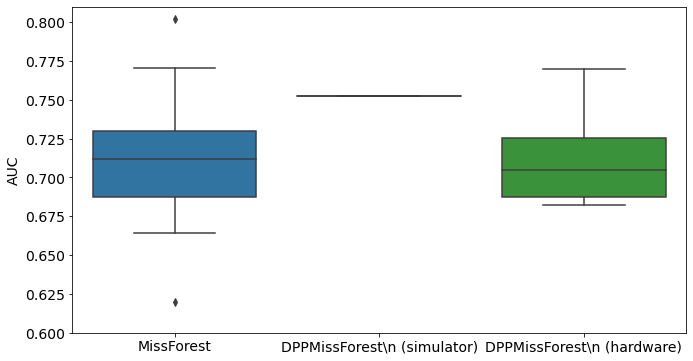}&\includegraphics[width=5.75cm]{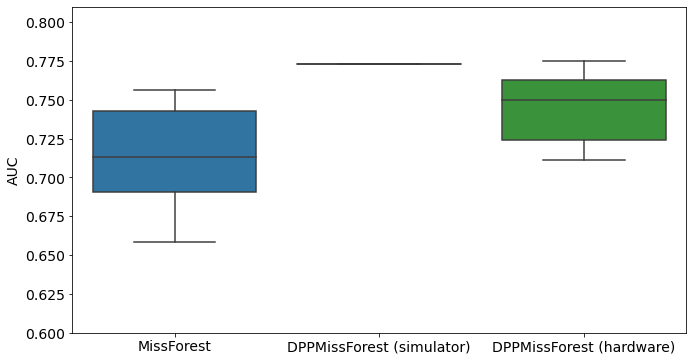}&\includegraphics[width=5.75cm]{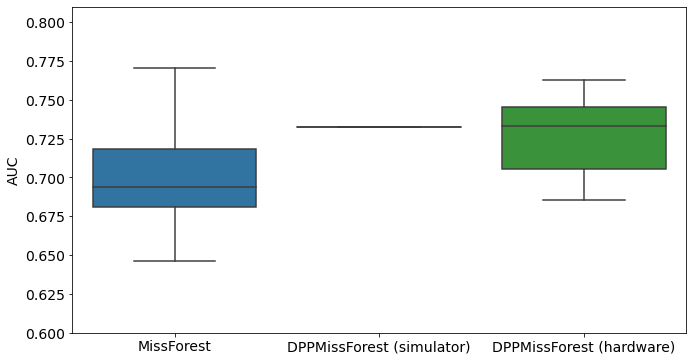}  \\
        \hline
        \rotatebox[y=1.5cm]{90}{MNAR  SYNTH}&\includegraphics[width=5.75cm]{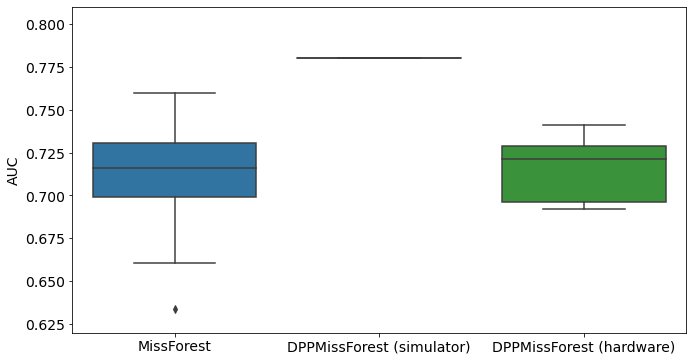}&\includegraphics[width=5.75cm]{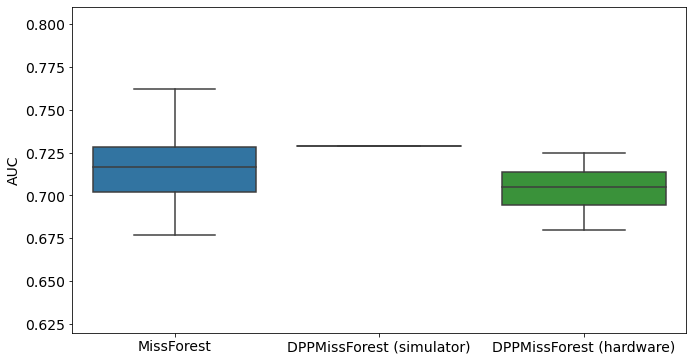}&\includegraphics[width=5.75cm]{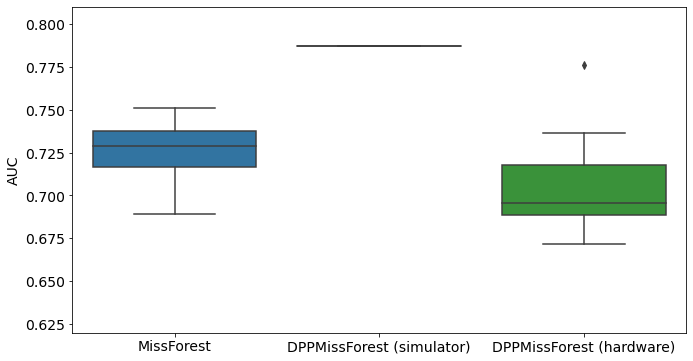} \\
        \hline
        \rotatebox[y=1.5cm]{90}{MNAR  MIMIC}&\includegraphics[width=5.75cm]{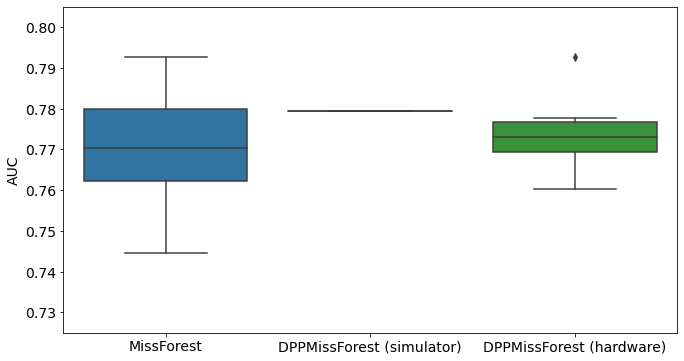}&\includegraphics[width=5.75cm]{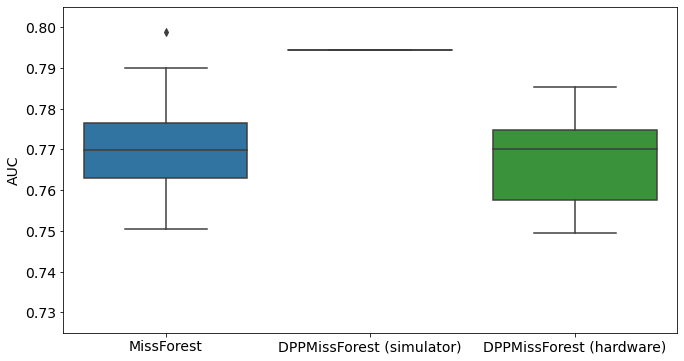}&\includegraphics[width=5.75cm]{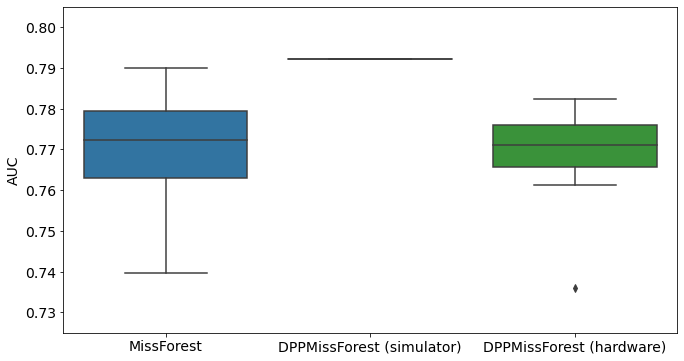}
         
    \end{tabular}
    \caption{Hardware results using the IBM quantum processor, depicting AUC results of the downstream classifier task after imputing missing values using DPP-MissForest. In the case of MissForest and the quantum hardware DPP-MissForest implementations, the boxplots correspond to 10 AUC values for 10 iterations of the same imputation and classification algorithm, depicting the lower and upper quartiles as well as the median of these 10 values. The AUC values are the same for every iteration of the quantum DPP-MissForest algorithm using the simulator.}
    \label{tab:hw_fig}
\end{table}

\begin{table}[!ht]
    \centering
    \begin{tabular}{|c|c|c|c|c|c|c|}
    \hline
        Dataset & Missingness & Batch size & Trees & MissForest & detDPP-MissForest (simulator) & detDPP-MissForest (hardware)  \\ \hline
        \multirow{6}{*}{SYNTH} & \multirow{3}{*}{MCAR} & 7 & 2 & 0.868±0.0302 & 0.9026 & 0.8598±0.021  \\  \cline{3-7}
        ~ & ~ & 8 & 3 & 0.8667±0.0342 & 0.9256 & 0.8923±0.027  \\  \cline{3-7}
        ~ & ~ & 10 & 4 & 0.8725±0.0275 & 0.9028 & 0.8902±0.024  \\  \cline{2-7}
        ~ & \multirow{3}{*}{MNAR} & 7 & 2 & 0.7122±0.0264 & 0.78 & 0.7149±0.02  \\ \cline{3-7}
        ~ & ~ & 8 & 3 & 0.7153±0.022 & 0.729 & 0.7036±0.0167  \\ \cline{3-7}
        ~ & ~ & 10 & 4 & 0.7258±0.0157 & 0.7868 & 0.7082±0.036  \\ \hline
        \multirow{6}{*}{MIMIC} & \multirow{3}{*}{MCAR} & 7 & 2 & 0.7127±0.038 & 0.7522 & 0.7117±0.0315  \\  \cline{3-7}
        ~ & ~ & 8 & 3 & 0.7136±0.03 & 0.7728 & 0.7448±0.0258  \\  \cline{3-7}
        ~ & ~ & 10 & 4 & 0.6968±0.03 & 0.7327 & 0.7262±0.0299  \\ \cline{2-7}
        ~ & \multirow{3}{*}{MNAR} & 7 & 2 & 0.7697±0.0133 & 0.7794 & 0.7742±0.0108  \\ \cline{3-7}
        ~ & ~ & 8 & 3 & 0.7713±0.0112 & 0.7943 & 0.767±0.0125  \\ \cline{3-7}
        ~ & ~ & 10 & 4 & 0.7712±0.0116 & 0.7922 & 0.7675±.01545 \\ \hline
    \end{tabular}
    \caption{Numerical quantum hardware results showing the AUC results of the downstream classifier task on reduced datasets. Values are represented according to Mean±SD (standard deviation) format given 10 values for each experiment.}
    \label{tab:hw}
\end{table}

\section*{DISCUSSION}

Missing data is a critical issue for machine learning practitioners as complete data sets are usually required for training machine learning algorithms. To achieve complete data sets, missing values are usually imputed. In the case of clinical data, missing values and imputation can be a potential source of bias and can considerably influence the robustness and interpretability of results. Nevertheless, there is no canonical way to deal with missing data which makes improvements in data imputation methods an attractive and impactful approach to increase the effectiveness and reliability of clinical trials. In this proof of concept study, we assessed the downstream consequences of implementing such improvements focussing on MCAR and MNAR to assess the usefulness of our approach. MNAR and MCAR represent two extreme cases of missingness with importance for clinical data imputation applications.

Determinantal point processing methods increase the diversity of the data picked to train the models, showcasing also that data gathering and pre-processing are important to remove biases related to over-representation of particular data types. This is more important when dealing with unbalanced datasets, as is the case often with clinical data. Determinantal sampling is an important tool not only for Random Forest models, but also for linear regression, where data diversity results in more robust and fair models \cite{D18}. Moreover, such sampling methods based on determinantal point processes are computationally intensive and quantum computers are expected to be useful in this case: quantum computers offer an asymptotic speedup for performing this sampling and it is expected that next generation quantum computers will provide a speedup in practice. 

We show that, as expected, the quantum version of detDPP-MissForest does not introduce any variance in the downstream classifier when simulated in the absence of hardware noise. While the AUC improvements achieved in our experiments may seem modest, it is the consistency of improvements we observed in our simulation results coupled with removal of variance that makes our approach attractive for clinical data applications where these characteristics are extremely desirable. When implemented on quantum hardware, we observed variance that is caused by the noise in the hardware itself. More precisely, the output of the quantum circuit is not a sample from the precise determinantal distribution but from a noisy version of it, and this noise depends on the particular quantum circuit implemented and the quality of the hardware. Thus when attempting to compute the highest probability element using samples from the quantum circuit on current hardware, the result is not deterministic. Importantly, unlike for standard MissForest, this variance is not inherent in the algorithm and is expected to reduce considerably with the advent of better quality quantum computers. The quantum circuits needed to efficiently perform determinantal sampling require a number of qubits equal to the batch size used for each decision tree within the Random Forest training and the depth of the quantum circuit is roughly proportional to the number of features. As an example, if we would like to perform the quantum version of the determinantal imputation methods we used for MIMIC-III, then we would need a quantum computer with 150 qubits (the batch size) that can be reliably used to perform a quantum circuit of depth around 400 (the depth is given by $4d\log n$, where $n=150$ is the batch size and $d=14$ is the number of features \cite{KP22}). While quantum hardware with a few hundred qubits that can perform computations of a few hundred steps are not available right now, it seems quite possible that they will be available in the not so far future. In the meantime further optimization could also help reduce the quantum resources needed for such imputation methods.

While our DPP-based imputation methods can run classically on small datasets such as MIMIC-III, they are computationally intensive and are hard to parallelize due to the sequential nature of the algorithm. This results in less and less efficient imputation for larger datasets where DPP sampling is applied to bigger batches. For example, when a DPP-MICE imputation is run on a dataset of 200 features and batches of size 400, then the training is expected to take multiple hours on a single GPU.   
The quantum DPP algorithm therefore could  provide a way to speed up the hardest part of the imputer using a next-generation quantum computer. For instance, if $d=200$, and batch size is 400, the number of qubits will be 400 and the depth of the quantum circuit would be $\approx 6400$, whereas it would take $\sim 8*10^6$ classical steps for DPP sampling. These are of course simply illustrative calculations and will require more detailed analysis as these machines become available and will need to include parameters such as clock speeds and error correction overheads. Only then can it be experimentally proven that this theoretical asymptotic speedup can translate to a practical speedup for this particular algorithm.

In summary, here we propose novel data imputation methods that: first, improve the widely-used iterative imputation methods –MiceRanger and MissForest– as measured by the AUC of a downstream classifier; second, remove the variance of the imputation methods, thus ensuring reproducibility and simpler integration into machine learning workflows; third, become even more efficient when run on quantum computers. Based on our results, we anticipate an impact of our algorithms on the reliability of models in high precision value settings, including in pharmaceutical drug trials where they can provide higher confidence in the predictions made by eradicating the stochastic variance due to multiple imputations. In addition, tasks that are currently overwhelmed by the challenges of missingness become more tractable through the approaches introduced here, which is a common problem with real-world-evidence investigations, where detDPP-MICE and detDPP-MissForest can yield improved performance in the face of missingness.

\section*{METHODS}

\subsubsection*{Determinantal Point Processes (DPPs)}

Given a set of items $\mathcal{Y} = \{y_1, \dots, y_N\}$, a point process $\mathcal{P}$ is a probability distribution over all subsets of the set $\mathcal{Y}$. It is called a Determinantal Point Process (DPP) if, for any subset $Y$ drawn from $\mathcal{Y}$ according to $\mathcal{P}$, we have:

\begin{equation}
    \mathcal{P}( T \subseteq Y) = \operatorname{det}(\bold{K}_{T,T}),
\end{equation}

where $\bold{K}$ is a real symmetric $N \times N$ matrix, and $K_{T,T}$ is its submatrix whose rows and columns are indexed by $T$. The matrix $\bold{K}$ is called the marginal kernel of $Y$.

For an $n \times d$ data matrix $A$ and $\bold{L}=AA^T$, we define the $\bold{L}$-ensemble $\mathrm{DPP}_L(\mathbf{L})$ as the distribution where the probability of sampling $T$ is:

\begin{equation}
    P(\{T\}) = \frac{\operatorname{det}(\bold{L}_{T,T})}{\operatorname{det}(\bold{I}+\bold{L})} \propto Vol^2(\{a_i: i\in T\}),
\end{equation}
where $Vol(\{a_i: i\in T\})$ is the volume of the parallelepiped spanned by the rows of $A$ indexed by $T$. 

According to this distribution, the probability of sampling points which are similar and thus form a smaller volume is reduced in favor of samples which are more diverse.

An $\bold{L}$-ensemble is a Determinantal Point Process if $\bold{K}=\bold{L}(\bold{I}+\bold{L})^{-1}$.

\hfill
\paragraph*{Stochastic $k$-DPPs.}

The distribution $k-\text{DPP}_L(\bold{L})$ is defined as an $\bold{L}$-ensemble which is constrained to subsets of size $|T|=k$.

Different algorithms have been proposed in the literature to sample from $k-\text{DPP}s$, namely \cite{KT13} where sampling $d$ rows from an $N\times d$ matrix takes $O(Nd^2)$ time. There have been improvements over this initial proposal as in \cite{DC19} where there is a preprocessing cost of $O(Nd^2)$ and each DPP sample requires $O(d^3)$ arithmetic operations.

\hfill

\paragraph*{Deterministic $k$-DPPs.} 

Stochastic DPP sampling may be efficient in practice, however deterministic algorithms are important for different use cases since they are more interpretable, are less prone to errors and have no failure probability, which is especially relevant for clinical data \cite{inter}. 

We use a deterministic version of DPP sampling as proposed in \cite{SF21} (see Algorithm 1) which is a greedy maximum volume approach. For each deterministic $k-\text{DPP}$ sample, elements with the highest probability are added iteratively. The complexity of the algorithm for selecting deterministically $d$ rows from a $N \times d$ matrix is $O(N^2 d)$ for the preprocessing step and $O(Nd^3)$ for the sampling step.

\begin{algorithm}[h]

\label{alg:1}
\KwIn{: $ N \times N$ Kernel matrix $K\succ 0$, sample size $k$.} 

\textbf{Initialization} : $\mathcal{T} \leftarrow \emptyset$

$\mathrm{V} \in \mathbb{R}^{n \times k}$: first $k$ eigenvectors of $K$.

$P=V V^{\mathrm{T}}$

$ p_0(i)=\left\|\mathrm{V}^T e_i\right\|^2, \quad i=1 \dots k$ 

$p \leftarrow p_0$ and $i=1$

\textbf{while} : $i \leq k$ \textbf{do}

\quad \quad $t_i \in \arg\max p$

\quad \quad $\mathcal{T} \leftarrow \mathcal{T} \cup\left\{t_i\right\}$

\quad \quad $p(j)=p_0(j)-P_{\mathcal{T} j}^T P_{\mathcal{T T}}^{\dagger} P_{\mathcal{T} j}, \quad j=1\dots n$

\quad \quad $i \leftarrow i+1$

\bf end while 

\KwOut{$\mathcal{T}.$}

\caption{Deterministic k-DPP algorithm}

\end{algorithm}

\subsubsection*{DPP-Random Forest and detDPP-Random Forest} \label{DPP-MICE}

The Random Forest is a widely-used ensemble learning model for classification and regression problems. It trains a number of decision trees on different samples from the dataset, and the final prediction of the Random Forest is the average of the decision trees for regression tasks or the class predicted by the most decision trees for classification tasks.

\begin{figure}[!h]
    \centering
    \includegraphics[width=0.6\linewidth]{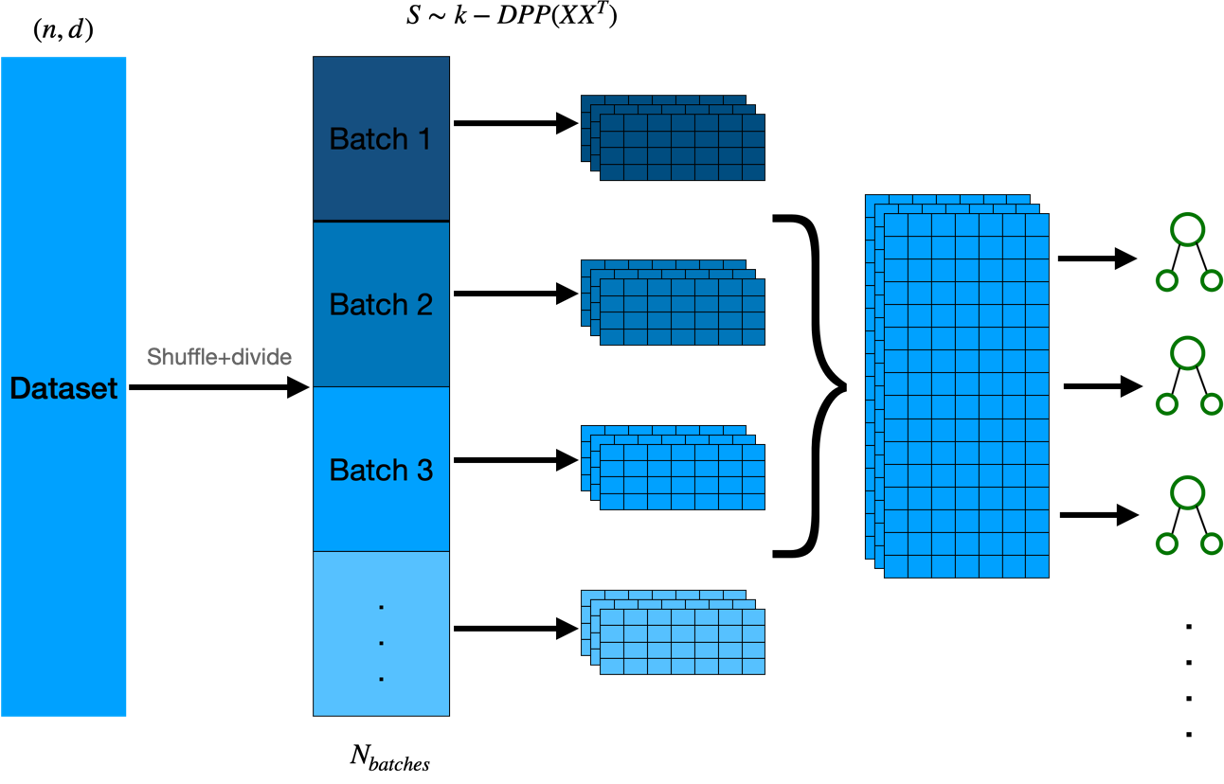}
    \caption{The sampling and training procedure for the DPP-Random Forest algorithm: the dataset is divided into batches of similar size, the DPP sampling algorithm is then applied to every batch in parallel, the subsequent samples are then combined to form larger datasets used to train the decision trees. Since the batches are fixed, DPP sampling can be easily parallelized, either classically or quantumly.}
    \label{fig:dpp_rf_method}
\end{figure}

The samples used to train each tree are drawn uniformly with replacement from the original dataset (bootstrapping). The DPP-Random Forest algorithm  (see Fig. \ref{fig:dpp_rf_method}) replaces the uniform sampling with DPP sampling without replacement. 

The running time of the standard Random Forest training on a $N \times d$ matrix is $\Tilde{O}(Nd)$, whereas the DPP-Random Forest algorithm takes $\Tilde{O}(Nd^2 + d^3)$ steps to run. This shows that while for small $d$ the classical DPP-enhanced algorithms can still be efficient, they quickly become inefficient for larger feature spaces.

Determinantal sampling for regression and classification tasks with full data has been proposed previously for Linear Regressors \cite{D18} and for Random Forest training for a financial data classification use case where it outperformed the standard Random Forest model \cite{itau23}.

\begin{figure}
    \centering
    \includegraphics[width=0.55\linewidth]{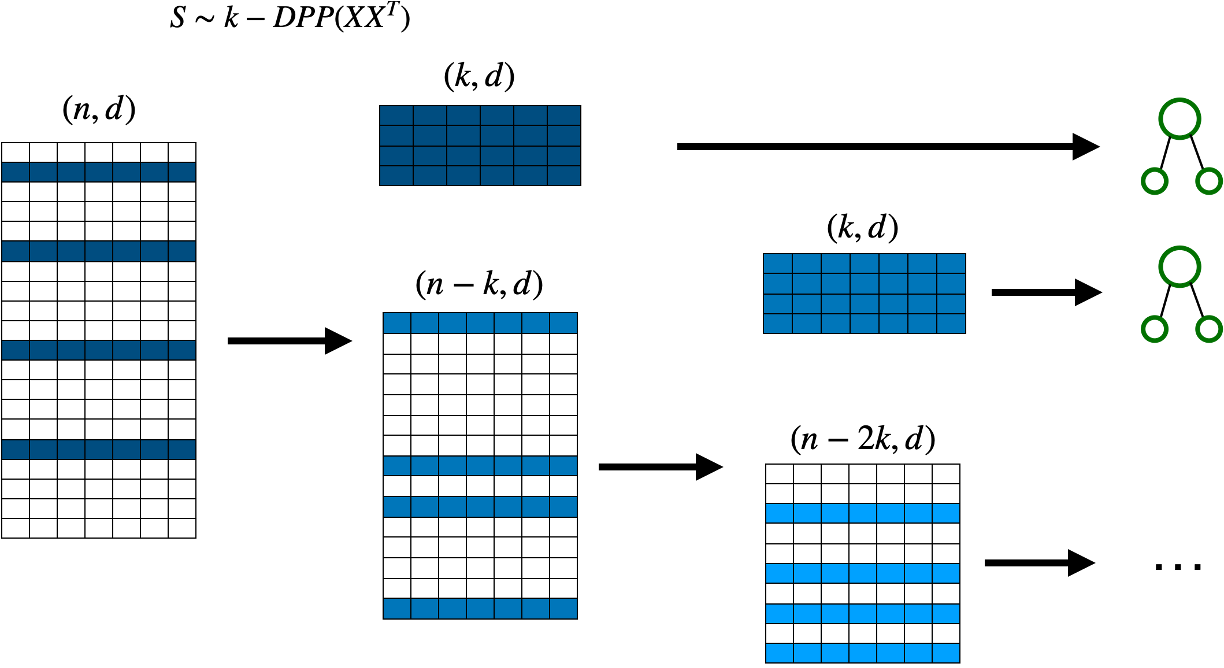}
    \caption{Deterministic DPP sampling procedure for training decision trees. At each step, a decision tree is trained usingthe sample that corresponds to the highest determinantal probability, and which is then removed from the original batch before continuing to the next decision tree.}
    \label{fig:def_dpp_rf_method}
\end{figure}

We can also use the deterministic version of DPP sampling for the Random Forest algorithm. This requires removing the sample used at each step (which is the one with the highest probability according to the determinantal distribution) in order to create a smaller dataset from which to sample for the next decision tree (see Fig. \ref{fig:def_dpp_rf_method}). We call this new model detDPP-Random Forest.

Let us note that the distributions of the in-bag DPP samples, which are biased towards diversity, and the out-of-bag (OOB) samples, which reflect the original dataset's distribution, may be different. This could lead to an inaccurate calculation of the OOB error that can be in fact overestimated \cite{oob}. In the DPP-Random Forest case, the batches are stratified and according to the output variable that follows the same distribution as the original dataset. Thus, sampling from different batches could bridge the gap between the in-bag and the out-of-bag distributions. We leave these considerations for future work.

\subsubsection*{Quantum methods for  DPPs}

Quantum Machine Learning has been a rapidly developing field and many applications have been explored, including with biomedical data, both using quantum algorithms to speedup linear algebraic procedures and through quantum neural networks \cite{Cerezo2022, Biamonte2017, Landman2022quantummethods, vision2022}. 

In \cite{KP22}, it was shown that there exist quantum algorithms for performing the determinantal sampling with better computational complexity than the best known classical methods. We describe below the quantum circuits that are needed for performing this quantum algorithm on quantum hardware with different connectivity characteristics and provide a resource analysis for the number of qubits, the number of gates and the depth of the quantum circuit.


First, we introduce an important component of the quantum DPP circuit which is the Clifford loader. Given an input state $x \in \R^n$, it performs the following operation:
$$\mathcal{C}(x) = \sum_{i=1}^n x_i Z^{i-1} X I^{n-i}$$ In other words it encodes the vector $x$ as a sum of the mutually anti-commuting operators generating the Clifford algebra.

For implementing this operation with an efficient quantum circuit, we use standard one- and two-qubit gates, such as the X, Z, CZ gates as well as a parameterized two-qubit gate called the Reconfigurable Beam Splitter gate (RBS), which does the following operation:

\begin{equation} \label{RBS}
RBS(\theta) = \left( \begin{array}{cccc}
1 & 0 & 0 & 0 \\
0 & \cos \theta & \sin \theta & 0 \\
0 & -\sin\theta & \cos\theta & 0 \\
0 & 0 & 0 & 1  \end{array} \right)
\end{equation} 

\begin{figure}
\centering
\begin{subfigure}{.3\textwidth}
    \centering
    \includegraphics[width=\linewidth]{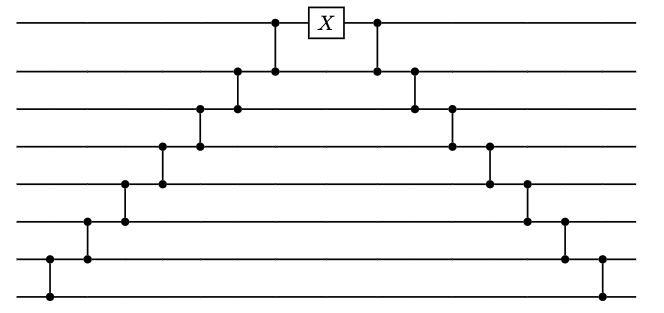}
    \caption{Diagonal Clifford loader}
    \label{fig:diag_cliff}
\end{subfigure}%
\begin{subfigure}{.3\textwidth}
    \centering
    \includegraphics[width=0.8\linewidth]{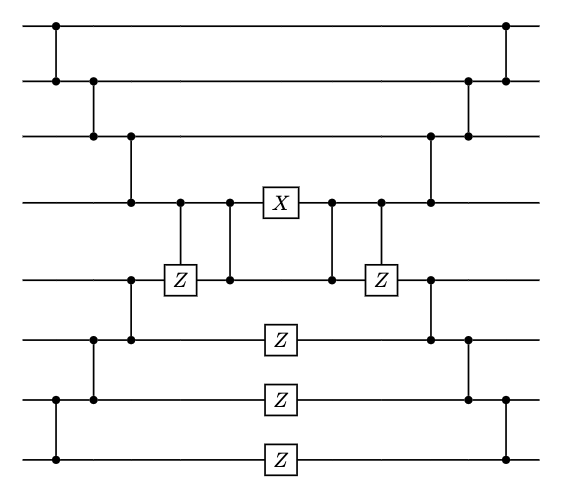}
    \caption{Semi-diagonal Clifford loader}
    \label{fig:semidiag_cliff}
\end{subfigure}%
\begin{subfigure}{.3\textwidth}
    \centering
    \includegraphics[width=\linewidth]{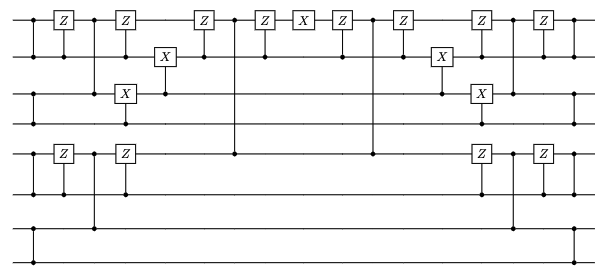}
    \caption{Parallel Clifford loader}
    \label{fig:parallel_cliff}
\end{subfigure}
    \caption{Types of data loaders. Each line corresponds to a qubit. Each vertical line connecting two qubits corresponds to an RBS gate. We also use $X,Z,CZ$ gates. The depth of the first two loaders is linear and the last one is logarithmic on the number of qubits.}
    \label{fig:loaders}
\end{figure}

We provide in Fig. \ref{fig:loaders} three different versions of the Clifford loader that take advantage of the specific connectivity of the quantum hardware, for example grid connectivity for superconducting qubits or all-to-all connectivity for trapped-ion qubits. These constructions are optimal (up to constant factor) on the number of two-qubit gates. We provide the exact resource analysis in Table \ref{tab:cliffloaders_tab}.

\begin{table}[!h]
    \centering
    \begin{tabular}{c|c|c|c}
        Clifford loader & Hardware connectivity & Depth & \# of RBS gates \\
        \hline 
        Diagonal & NN & $2nd$ &  $2nd$ \\
        \hline 
        Semi-Diagonal & NN & $nd$ & $2nd$ \\
        \hline 
        Parallel & All-to-all & $4d \log(n)$ & $2nd$
        
    \end{tabular}
    \caption{Summary of the characteristics of the different quantum DPP circuits. NN = Nearest Neighbor connectivity}
    \label{tab:cliffloaders_tab}
\end{table}

We can now use the Clifford loaders described above to perform $k-DPP$ sampling, as described \cite{KP22}. 

\begin{figure}[!h]
    \centering
    \includegraphics[width=0.5\linewidth]{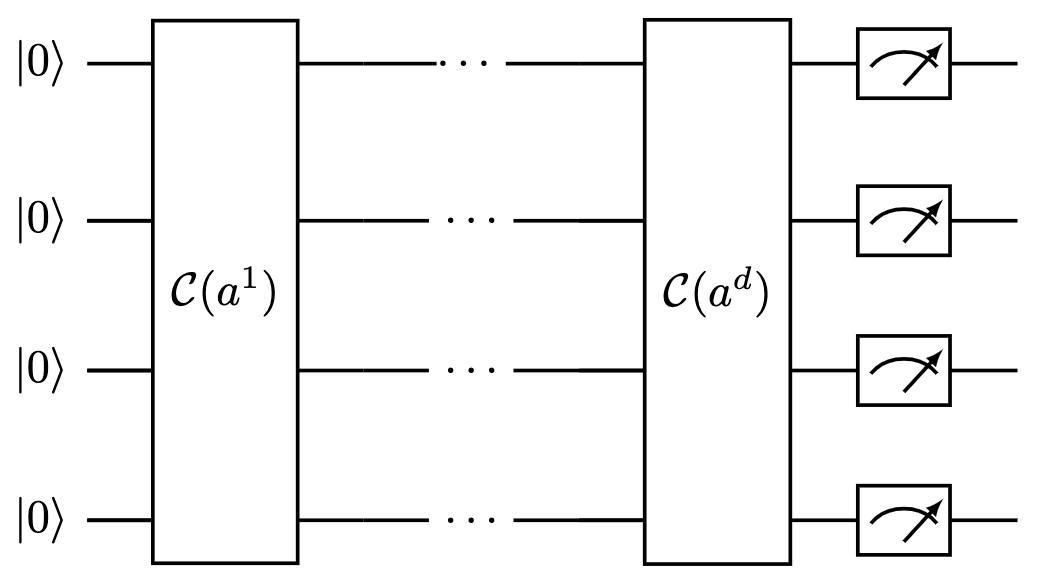}
    \caption{Quantum Determinant Sampling circuit for an orthogonal matrix $A=(a^1, ..., a^d)$. It uses the Clifford loader which is a unitary quantum operator: $\mathcal{C}(x) = \sum_{i=1}^n x_i Z^{i-1} X I^{n-i}, \quad \text{for}  \quad x \in \mathbb{R}^n$}
    \label{fig:qdpp_circuit}
\end{figure}

Given an orthogonal matrix $A=(a^1, ..., a^d)$, we can apply the qDPP circuit shown in Fig. \ref{fig:qdpp_circuit}, which is just a sequential application of $d$ Clifford loaders, one for each column of the matrix, to the $\ket{0^n}$ state, and that leads to the following result:

\begin{equation*}
    |\mathcal{A}\rangle=\mathcal{C}\left(a^d\right) \cdots \mathcal{C}\left(a^2\right) \mathcal{C}\left(a^1\right)\left|0^n\right\rangle = \sum_{|S|=d} \operatorname{det}\left(A_S\right)\left|e_S\right\rangle
\end{equation*}

Directly measuring at the end of the circuit provides a sample from the correct determinantal distribution. 

Both the classical and the quantum algorithms require a preprocessing step with a similar complexity (see Table \ref{tab:dpp_complexity}), the improvement using the quantum method achieves a quadratic to cubic speedup in the sampling step. This speedup holds for $n=O(d)$. This is the case for our current implementation of DPP sampling from smaller batches (see Fig. \ref{fig:dpp_rf_method}). In addition, the quantum DPP algorithm is efficient in terms of the number of measurements required since one measurement is equivalent to generating one DPP sample.

\begin{table}[!h]
    \centering
    \begin{tabular}{c|c|c}
        &Classical & Quantum \\\hline
        Preprocessing & $O(d^3)$ & $O(d^3)$ \\\hline
        Sampling & $\Tilde{O}(d^3)$ & $\Tilde{O}(d)$ depth \\ 
        &&$\Tilde{O}(d^2)$ gates\\
    \end{tabular}
    \caption{Complexity comparison of d-DPP sampling algorithms, both classical \cite{DC19} and quantum \cite{KP22}. The problem considered is DPP sampling of $d$ rows from an $n\times d$ matrix where $n=O(d)$. For the quantum case we provide both the depth and the size of the circuits.}
    \label{tab:dpp_complexity}
\end{table}

\paragraph*{Quantum versions of the imputation methods}

It is easy to define now a quantum version of the DPP-MICE and DPP-MissForest algorithms, where we use the quantum circuit described above to sample from the corresponding DPP. We can also define a variant of the deterministic algorithms, though here we need to pay attention to the fact that the quantum circuit enables to sample from the determinantal distribution but does not efficiently give us a classical description of the entire distribution. Hence one can instead sample many times from the quantum circuit and output the most frequent element. This provides a sample with less variance but it only becomes deterministic in the limit of infinite measurements. In the experiments we performed, we used 1000 shots and the samples from the quantum circuits were indeed most of the time the highest probability elements. Of course in the worst case, there exist distributions where for example the highest and second highest elements are exponentially close to each other, in which case the quantum algorithm would need an exponential number of samples to output the highest element with high probability. Note though that the quantum imputation algorithm will still have a good performance even with few samples (any high probability element provides the needed diversity of the inputs),  though it will not be deterministic.

\subsection*{Availability of Data and Code}

The code for the different DPP imputation methods is publicly available here \href{https://github.com/AstraZeneca/dpp_imp/}{github.com/AstraZeneca/dpp\_imp}.
The synthetic dataset can be generated using the \emph{make\_classification} method from scikit-learn. The MIMIC-III dataset \cite{MIMIC_Ref} is also publicly available.

\section*{Acknowledgements}

This work is a collaboration between QC Ware and
AstraZeneca. We acknowledge the use of IBM Quantum services for this work. The views expressed are those of the authors, and do not reflect the official policy or position of IBM or the IBM Quantum team.

\bibliographystyle{plain}
\bibliography{references}

\end{document}